\documentstyle[12pt]{article}
\documentstyle{fleqn}
\begin{document}
\begin{flushright}
OS-GE 33-93
\linebreak
July 17, 1993
\end{flushright}
\vskip0.3cm
\begin{center}
\large {\bf    Twisting Toroidal Magnetic Fields and
the Seasonal
Oscillation  of the Solar Neutrino Flux II
}
\end{center}
\vskip.3cm
\begin{center}
Takahiro Kubota, Takeshi  Kurimoto and  Eiichi Takasugi
\linebreak
\vskip.3cm
{\it Institute of Physics, College of General
Education,}
\linebreak
{\it Osaka University,  Toyonaka, Osaka 560, Japan
}
\end{center}
\vskip.3cm
\begin{abstract}
An intriguing possibility is explored   that the
solar neutrino data could be used as a probe to the
magnetic field structure in the sun.  Various cyclic
phenomena occurring on the surface of the sun have
been accounted for  by the so-called dynamo-mechanism.
According to this, a self-generating mechanism of
solar cyclic activities gives rise to twisting toroidal
magnetic fields in the convective zone. Although
its magnitude is not known,
the orientation  of the twist is certainly opposite in
northern and southern hemispheres of the sun.  We show
by numerical calculation that the solar neutrino flux,
being  sensitive to the twist, could exhibit
observable seasonal oscillation, provided that the
twist is sizable in magnitude and the neutrinos have
reasonably large magnetic moments.  This oscillation is
ascribed to the fact that the earth's orbit is
slightly inclined to the sun's equator, and that
solar neutrinos pass
through the northern (southern) hemisphere of the
sun around September (March).
We also argue that similar seasonal oscillation could
be exhibited  in the azimuthal asymmetry
of  recoiling
electrons scattered by the solar neutrinos which
are expected to be observed in the super-Kamiokande
detector.
\end{abstract}
\pagebreak
\begin{center}
{\bf 1. INTRODUCTION}
\end{center}
\vskip0.5cm

The deficit of the solar neutrinos observed in
Homestake [1],  Kamiokande II (KII) [2] and
the gallium experiments [3,4] has been one of the
strongest stimuli for theoretical ideas in elementary
particle physics.   Among many others,  the most
interesting mechanisms which attracted a lot of
attention  to solve the solar neutrino puzzles are :
(1)
resonant amplification of neutrino
oscillation proposed by Mikheyev and Smirnov and
Wolfenstein (MSW) [5],  and
(2)
 possibility of large  magnetic moments of
neutrinos,  which  convert solar neutrinos into sterile
right-handed ones [6,7].
These seminal  ideas were followed by
further elaboration thereof [8] and a few review
articles [9] on these  subjects are now available.

These mechanisms  have been expected to  open
 new windows in  neutrino physics, and to suggest
what lies  beyond the standard model.  In the present
paper, however,    we would like to take
a little bit different
attitude of looking at the solar neutrino data:
we will explore  a possibility to make use of the solar
neutrino data to   investigate  the magnetic structure of
the sun, supposing that neutrinos are endowed with
reasonably large magnetic moment.

The observation of solar activities has a long history of
its own. Everybody now agrees that phenomena occurring
on the surface of the sun are outer  manifestations of
inner dynamics which controls the magnetic structure of
the so-called convective zones of the sun.  A task set
upon solar physicists has been to find a coherent
explanation of  self-generating mechanisms  of the
cyclic phenomena of the sun. The cycle is such that
the toroidal  magnetic
force lines  undergo periodical change in orientation
in the convective zones, which leads to the 11-year
cycle of the sunspots occurrence.

One of the  important implications of the
solar dynamo is that the toroidal magnetic field receives
twisting in the course of its time-development.
Interesting enough is that the orientation of the
twisting is very likely to be opposite in the northern
and southern hemispheres of the sun.  We will focus our
eyes on this peculiar fact, and will  investigate the
possibility that the evidence in favor of
opposite twisting
in each hemisphere has already been encoded in  the solar
neutrino data.

The study of the neutrino propagation in the presence of
twisting magnetic fields was started by Vidal and Wudka
[10] and also by Aneziris and Schechter [11].
Smirnov [12] later clarified the implications
of the twist and pointed out a new mechanism of resonant
neutrino    oscillations.
Qualitatively we can easily speculate the following
phenomena. First of all let us recall that the earths's
orbit makes an angle of $7.25 ^{\circ}$ with the plane
of the solar equator.
In other words, neutrinos observed on the earth have
passed through the northern (southern) hemisphere
of the sun around September  (March).  This means
that neutrinos would experience twisting magnetic
fields in opposite directions depending on the
season, {\it i.e.,} effects of the twisting fields
on the neutrino flux would exhibit semi-annual
variation.

The purpose of the present paper is to put these
qualitative considerations  on a more quantitative
basis.  Preliminary results have been reported
in ref.  [13].   We will show here that, with
reasonably large magnetic moments,  solar magnetic
fields  and sizable twisting,  the total amount of
solar neutrino flux should go up and down every
year.  We will also
argue that there should be seasonal change in the
observation of the azimuthal asymmetry in the
distribution  of recoiling electrons
scattered by neutrinos  which are to be observed
in the super-Kamiokande  detectors.

The seasonal variation  to be discussed in this paper
reminds us of  the similar one discussed several years
ago by Okun, Voloshin, and Vysotsky [6].
They were  aware of the fact that
there is a small equatorial ``slit" in the solar magnetic
field where the magnetic strength is very weak.   They
noticed  that there would be a maximum in the
left-handed neutrino flux in  June and December.
Perhaps it should be born in our mind  that, while their
ideas are interesting in their own right,  the
origin of their seasonal variation differs from
ours. In their case the latitude dependence of the
magnetic field strength is the source of the variation,
while in our case   the twist is playing the
crucial role.

The present paper is organized  as follows. First of all
in sec. 2,
 we recapitulate the neutrino evolution equation to explain
how the twisting magnetic field could affect the neutrino
flux.  In sec. 3, we describe the profile of the solar
magnetic field,
and set up a simple model for it which will be used in
our numerical calculations in sec. 5.
The azimuthal asymmetry of electron recoil  in KII
(and/or super-Kamiokamde ) detectors
is discussed in sec. 4 on the basis of the model in sec. 3.
Sec.  6 is devoted to discussions and summary.
\vskip1cm
\begin{center}
{\bf 2. EVOLUTION EQUATION AND TWIST}
\end{center}

Let us begin with the evolution equation of neutrinos
in the presence of twisting magnetic fields.
The evolution equation is usually discussed in a
parallel way with the neutral kaon system. In the
presence of  magnetic fields coupled to  neutrino
spin,  the evolution equation is
to be derived on the
basis of relativistic wave equations.
For a pedagogical derivation, readers are referred to
Mannheim's paper [14], where both Dirac and Majorana
cases are discussed.

In our present investigation, we consider Majorana
neutrinos, whose time evolution is described by
the Schr\"odinger equation
\begin{equation}
i\frac{d\Psi }{dt}=H\Psi,
\end{equation}
where we include into our state vector $\Psi$,
 electron and muon
neutrinos and their charge conjugation  (denoted
by $C$)
\begin{equation}
\Psi ^{T}=\bigl (\nu _{eL}, \nu _{\mu L}, (\nu _{eL}
)^{C}, (\nu _{\mu L})^{C}\bigr ).
\end{equation}
Inclusion of tau neutrinos are rather straightforward.
To give a definite form of the Hamiltonian in (1),
we take henceforth the moving direction of neutrinos
the $z$-axis. Thereby the Hamiltonian becomes.
\begin{equation}
H=\left( \begin{array}{cc}
             H_{+} &   M \\
             M^{\dag} &  H_{-} \\
   \end{array}
   \right),
\end{equation}
where
\begin{equation}
H_{\pm}=\left (\begin{array}{cc}
                \pm V_{e} &  \Delta m^{2}
           \sin \theta /4E \\
           \Delta m^{2}\sin \theta /4E &
           \pm V_{\mu }+\Delta m^{2}
           \cos 2\theta /2E \\
\end{array}
\right),
\end{equation}
\begin{equation}
M=\left( \begin{array}{cc}
           0 &  -\mu (B_{x}+iB_{y}) \\
           -\mu (B_{x}-iB_{y}) & 0 \\
\end{array}
\right).
\end{equation}
The matter potentials perceived by electron and muon
neutrinos are denoted by $V_{e}=G(2N_{e}-N_{n})/
\sqrt{2}$ and $V_{\mu}=-GN_{n}/\sqrt{2}$,
respectively,
where $N_{e}$ and $N_{n}$ are the number densities of
electrons and neutrons, respectively.
The lepton-number violating Majorana masses $m_{ee}$,
 $m_{\mu \mu}$ and  $m_{e\mu}$ appear in the evolution
equation only in  the
combination of
$
\Delta m^{2}=\sqrt{(m_{ee}^{2}-m_{\mu \mu}^{2})^{2}+
4m_{e\mu }^{4}},
$
 together with the mixing angle $\theta $.

In the course of deriving eq. (3), it has been assumed
that the neutrino energy $E$, neutrino (transition)
magnetic moment $\mu$, magnetic field strength $B$,
and   neutrino masses satisfy $E\gg m_{ij}
 (i,j=e, \mu ), E\gg\mu  B$.
The reason for the presence of the term
\begin{equation}
B_{x}\pm iB_{y}\equiv B_{T}\exp (\pm i\phi)
\end{equation}
is due to the fact that the off-diagonal elements
 in $M$ come from the spin-fliping terms of the
magnetic moment  interactions.  Terms that contain
the $z$-component of the magnetic field $B_{z}$ would
appear in the off-diagonal part of $H_{\pm}$.
They are, however, suppressed by a power $\mu B_{z}/E$
and have been neglected in (3) in our approximation.
The phase $\phi $ is often referred to as geometrical
 or  topological  phases, because it would give rise to
Berry's phase [15] if it were adiabatically
time-dependent.

To have a physical insight into the
phase $\phi $
in eq. (6),  it is more sensible to truncate  our
$4\times 4$ matrix  into the following
$2\times 2$ matrix equation
\begin{equation}
i\frac{d}{dz}
\left( \begin{array}{c}
                     \nu _{eL} \\
                     (\nu _{\mu L})^{C}\\
\end{array}
\right)
=\left (\begin{array}{cc}
             V/2  & -\mu B_{T}\exp (i\phi ) \\
             -\mu B_{T}\exp (-i\phi ) & -V/2 \\
\end{array}
\right )
\left( \begin{array}{c}
                     \nu _{eL} \\
                     (\nu _{\mu L})^{C}\\
\end{array}
\right).
\end{equation}
The potentials $V_{e}$ and $V_{\mu }$ have been replaced
in the above by $\pm V/2$, where
\begin{equation}
V\equiv V_{e}+V_{\mu }=\sqrt{2}G(N_{e}-N_{n}).
\end{equation}
This is allowed because we can always take out common
diagonal phase.  Note also that the time derivative
has been replaced by the $z$-derivative in (7), since
neutrinos travel nearly with the light velocity.
The mass term $\Delta m^{2}$ has been neglected in (7) in
order to make our argument as definite as possible.
The combination of $\nu _{eL}$ and $(\nu _{\mu L})^{C}$
 as a Dirac field has been considered by Zel'dovich
and Konopinski and Mahmoud (ZKM) [16]  and will be
referred to as ZKM Dirac scheme.

The effect of geometrical phase was first discussed
in the hope that the existence of the phase might
be efficient enough to solve the solar neutrino problem
even with very small neutrino magnetic moment.  Further
investigation, however, showed that this hope was an
illusion  and that the adiabatic approximation used
often in connection with  Berry's phase was not
applicable in this problem [11, 12, 17].  Although
the original hope was turned down, Smirnov [12] realized
that there exists a potentially interesting conspiracy
between the geometrical phase and the resonant spin-flip
as well as resonant spin-flavor transitions.

The best way to see the effect of the
phase is, following Smirnov,  to transfer to the new
 state basis
\begin{equation}
\left( \begin{array}{c}
                     \tilde{\nu} _{eL} \\
                     (\tilde{\nu} _{\mu L})^{C}\\
\end{array}
\right)
=
\left (
\begin{array}{cc}
\exp (-i\phi /2) & 0 \\
0 & \exp (i \phi /2) \\
\end{array}
\right )
\left( \begin{array}{c}
                     \nu _{eL} \\
                     (\nu _{\mu L})^{C}\\
\end{array}
\right ).
\end{equation}
The change of the basis transforms eq. (7) into
\begin{equation}
i\frac{d}{dz}
\left( \begin{array}{c}
                     \tilde{\nu} _{eL} \\
                     (\tilde{\nu} _{\mu L})^{C}\\
\end{array}
\right)
=\left (\begin{array}{cc}
             (V+\phi ')/2  & -\mu B_{T} \\
             -\mu B_{T} & -(V+\phi ')/2 \\
\end{array}
\right )
\left( \begin{array}{c}
                     \tilde{\nu} _{eL} \\
                     (\tilde{\nu} _{\mu L})^{C}\\
\end{array}
\right),
\end{equation}
where $\phi '$ is the space derivative along  the
direction of the neutrino path.
It is now obvious from eq. (10) that spatial variation
of the phase
amounts to either increasing  ($\phi '>0$) or
decreasing ($\phi '<0$) the potential.
The resonant spin-flavor transition is modified
sensitively in accordance with the sign of
 $\phi '$.
\vskip1cm
\begin{center}
{\bf 3.  A SIMPLE MODEL OF TWISTING TOROIDAL
\linebreak
MAGNETIC FIELDS IN THE SUN}
\end{center}

The solar activities and in particular the solar
magnetic field profiles seem to be  extremely
complicated at first sight.  Observations over
many years,  however, revealed characteristic
patterns of solar activities. One of the most
familiar  facts is the cyclic occurrence
of sunspots in every 11 years. Observational
investigations show that the sunspot zones drift
equatorially  both in northern and southern
hemispheres. The magnetic nature of sunspots has
also been studied.  The sunspots
show up always in pair and the polarity of the pair
is opposite.
Notably, alternations in magnetic-field
polarity of bipolar sunspot groups in every 11 years
 have been confirmed and the oscillation period
is now doubled to 22 years.  The sunspots give us much
information of the lower and middle  latitude
of the sun, while around the poles the polarity
of the magnetic fields
are known to change at the maximum phase of the
solar cycles.

These observations were extremely  strong driving
forces to figure out
 the origin of the self-sustaining mechanism
of the cyclic phenomena.  An important ingredient to
study the solar magnetic fields is the
differential rotation of the sun. In plasma fluid
with high electric conductivities,  magnetic force lines
 tend to go along  the motion of the fluid. The
differential rotation  elucidates  therefore how to
generate toroidal magnetic fields from poloidal ones.
Although the differential rotation is crucial to
explain some of the solar phenomena,  another  mechanism
converting  the orientation of  the toroidal magnetic fields
 had to be looked for elsewhere
 to explain the cyclic
phenomena.

The global convection current has been called for
to fill in the missing link of the cycle.
 The global convection
is a global-scale fluid motion in the convective zone,
extending over the sphere.
Qualitatively, the Coriolis force is responsible
for twisting the magnetic fields and the time-evolution
processes of magnetic fields are illustrated in Fig. 1
schematically following the work of Yoshimura [18].
One can see in Fig. 1 that the toroidal
magnetic fields are twisted in the course of
development.  The reversed toroidal magnetic
fields locate deep inside the convective zone.  The
toroidal fields in the outer area are pushed toward
the surface of the sun, spread and create sunspots,
and then dissipate.  Eventually they are replaced
by the   reversed toroidal fields, that is
the polarity reversal at the maximum phase of the
solar cycle.

A quantitative analysis of the solar cycle requires
detailed study of the magneto-hydrodynamial (MHD)
equations, which describe the time evolution of
magnetic fields  under given   velocity
fields of plasma media.  To handle the MHD equations
analytically, however, is nearly impossible and
numerical integrations are the best  means to solve
the MHD equations.

Yoshimura [18] has been engaged in elaborating
numerical solutions to the MHD equations.
He prescribed the velocity fields of differential
rotation and global convections by introducing
several adjustable parameters.  His numerical
computations reproduce  the basic characteristics
of the solar activities.  It has been observed that
magnetic field tubes (torus) encircling the rotational
axis of the sun appear   naturally. The
 reversal of orientation of  toroidal magnetic fields
 takes  place inside the convection
zone.  The reversed magnetic fields propagate towards
the surface of the sun, which turns out to be solar
activities of the sunspots. The parameter dependence
of his simulation was also examined.  Although the
magnetic fields around the circling tube is twisting,
the twisting parameter is not uniquely determined.

In the present paper, being motivated by Yoshimura's
computer simulation, we will use a simplified model
of the    solar magnetic fields.  We consider
toroidal magnetic fields winding along the tori
which sit in the northern and southern convective areas
(Fig. 2).  Each torus is parametrized by its radius $a$,
the latitude of its center $\Delta$, and the distance
$b$ measured from the center of the sun.  The latitude
$\lambda$ in Fig. 2 shows the direction of the path
of neutrinos which are assumed to have been produced
at the center of the sun.

We take the magnetic fields to be distributed
symmetrically around   the rotating axis of the sun and
the twisting proceeds at the constant rate  along the
torus.  We introduce a parameter $X$ which parametrizes
the degree of the twist and is defined as the  distance
along the torus to wind it once. (See Fig. 3.)
The orientation of the twist varies in the first and
second 11 years and is  opposite in the northern and
southern tori. The patterns are schematically shown
in Fig. 4.

The direction of neutrino propagation is  taken to be $z$-axis
and the longitudinal (latitudinal)  direction
along the $x$ ($y$)-axis.
In this coordinate system the twisting
magnetic fields are parametrized by
\begin{equation}
\left(
\begin{array}{c}
B_{x}\\
B_{y}\\
B_{z}
\end{array}
\right)
=B
\left(
\begin{array}{c}
n_{x}\cos \theta (z)\\
n_{y}\sin \theta (z)\cos \alpha (z)\\
n_{z}\sin \theta (z)\sin \alpha (z)
\end{array}
\right).
\end{equation}
Here angles $\theta (z)$ and $\alpha (z)$ shown in
Fig. 5 are defined by
\begin{equation}
\theta (z)=\tan ^{-1}\big [\frac{2\pi R(z)}{X}\big ],
\end{equation}
\begin{equation}
\cos \alpha (z)=\frac{b\cos (\Delta-\lambda)-z}{R(z)},
\end{equation}
with
\begin{equation}
R(z)=\sqrt{[b\sin (\Delta -\lambda )]^{2}+[b\cos (\Delta
-\lambda )-z]^{2}}.
\end{equation}
The parameters $n_{x}$, $n_{y}$ and $n_{z}$ in eq. (11)
are sign factors specifying the twist which depends on
seasons and either  the
first or the second periods of 11 years.   They are
given from Fig. 4 by
\begin{equation}
(n_{x}, n_{y}, n_{z})=\left \{
\begin{array}{ll}
(+, +, -) & \mbox{for NI: northern hemisphere, first
11 yr,} \\
(-, +, +) & \mbox{for SI: southern hemisphere, first
11 yr,} \\
(-, -, +) & \mbox{for NII: northern hemisphere, second
11 yr,} \\
(+, -, -) & \mbox{for SII: southern hemisphere, second
11 yr.} \\
\end{array}
\right.
\end{equation}

So much for the parametrization of the magnetic field.
We are now in a position to evaluate $B_{T}$ and  the
phase $\phi $ from eq. (11), and we find
\begin{equation}
B_{T}=B\sqrt{\cos ^{2}\theta (z)+\sin ^{2}\theta (z)
\cos ^{2}\alpha (z)},
\end{equation}
\begin{equation}
\phi (z)=n\tan ^{-1}[\frac{2\pi}{X}(b\cos (\Delta
-\lambda)-z)]+\delta .
\end{equation}
Here a sign factor $n$ and $\delta $ are given by
$(n,\delta)=(+1,0),$  $(-1, \pi ),$   $(+1, \pi ),$
 $(-1, 0)$ for NI,  SI,  NII,  and SII, respectively.
We are interested in the $z$-derivative of $\phi$
in the evolution equation
\begin{equation}
\phi '(z)=
\left \{\begin{array}{c}
-1 \\
1
\end{array}
\right \}
\frac{2\pi/X}{1+(2\pi /X)^{2}[b\cos (\Delta -\lambda)-z]^{2}}
{}~~~~{\rm for}
\left \{
\begin{array}{c}
{\rm NI, NII,} \\
{\rm SI, SII.}
\end{array}
\right.
\end{equation}
This shows clearly  that the effect of the phase depends
crucially on whether neutrinos pass through the northern
or southern hemispheres of the sun.

Qualitatively, if
neutrinos go through the northern hemisphere (around
September), then the negative sign on the right hand
side of eq. (18)
 indicates suppression of the
the diagonal element of the Hamiltonian in (3), resulting in
an relative enhancement of the off-diagonal effect.
We can expect  large spin-flip  conversion rate around
September and considerable  neutrino flux suppression.
On the contrary if neutrinos are passing through southern
hemisphere (around March), then the neutrino flux will
be less suppressed.
The seasonal contrast depends on the magnitude of the
magnetic field.
If the magnetic field strength is larger (e.g., in the
maximum phase    of the solar cycle),  then the seasonal
variation is rather veiled.  This is qualitatively
different from what Okun,  Voloshin and Vysotsky [6] have
considered. In their case, larger the magnetic field, more
conspicuous the seasonal variation.
\vskip1.0cm
\begin{center}
{\bf 4. AZIMUTHAL ASYMMETRY OF THE  ELECTRON RECOIL}
\end{center}

Let us turn to another possibility to detect the
twist of the solar magnetic field.   A conceivable
direct evidence of the twist could be found in
the azimuthal asymmetry in the distribution
of the electrons scattered by
solar neutrinos. The origin of the asymmetry is the
interference between
electromagnetic scatterings due to the neutrino
magnetic moment and weak processes. Such an asymmetry
has been discussed  by Barbieri and Fiorentini [19]
in their search for an evidence of helicity conversion.
They considered  Dirac neutrinos with magnetic
moment.

Neutrinos produced at the center of the sun are
purely in the left-handed state, whose probability
amplitude
is given by $(\tilde{\nu }_{e L},  (\tilde{\nu }_
{\mu L})^{C})=(1,0)$.
In the course of traversing the solar convective zones,
the amplitudes acquire complex phases.
Suppose that the
neutrinos  just entering the super-Kamiokande detector are
described by
the   amplitudes ~~
\begin{equation}
(\tilde{\nu }_{eL},  (\tilde{\nu }_{\mu L})^{C})
=(\mid a_{L} \mid
e^{i\varphi _{L}},   \mid a_{R} \mid e^{i\varphi _{R}})
{}.
\end{equation}
  The spin of the neutrinos then has an expectation
value
\begin{equation}
<{\bf S}>=\left (
\begin{array}{c}
\mid a_{L}a_{R}\mid \cos (\varphi _{L}-\varphi _{R}) \\
\mid a_{L}a_{R}\mid \sin (\varphi _{L}-\varphi _{R}) \\
(\mid a_{R}\mid ^{2}-\mid a_{L}\mid ^{2})/2.
\end{array}
\right ).
\end{equation}
The phases $\varphi _{L}$, and $\varphi _{R}$ of
the left and right components are calculable
numerically by use of (10).  We immediately notice
that the particular direction of the neutrino spin (20)
may produce an asymmetry with reference
to this direction.   In fact, an  asymmetry
can  be found  if we look at the  azimuthal
distribution of the recoiling  electrons.
The phases $\varphi _{L}$ and $\varphi _{R}$ are
sensitive to the twisting magnetic field in the sun,
we might be able to expect a semi-annual variation
in the azimuthal asymmetry data.

The differential cross sections relevant to our case
 becomes as follows:
\begin{equation}
\frac{d\sigma}{dTd\varphi}=\frac{d\sigma ^{\nu _{L}}_{W}}
{dTd\varphi}+\frac{d\sigma _{em}}{dTd\varphi}+\frac{d\sigma
^{\nu _{L}}_{int}}{dTd\varphi}+
\frac{d\sigma _{W}^{\bar \nu _{R}}}{dTd\varphi}+\frac{
d\sigma _{int}^{\bar \nu _{R}}}{dTd\varphi}.
\end{equation}
Here, $T$ is the electron kinetic energy and $\varphi $
is the azimuthal angle of the recoiling  electron.
The weak gauge-boson exchange
contributions for $e\nu _{eL} \rightarrow e\nu _{eL}$
($e( \nu_{\mu L})^{C} \rightarrow e(\nu _{\mu L})^{C}$)
 and electromagnetic one  are
denoted by $d\sigma
_{W}^{\nu _{L}}$ ($d\sigma _{W}^{{\bar \nu }_{R}}$)
 and $d\sigma _{em}$, respectively.  These cross
sections together with
 their interference  ( $d\sigma _{int}^{\nu _{L}}$
and $d\sigma _{int}^{{\bar \nu }_{R}}$)
are explicitly given by
\begin{equation}
\frac{d\sigma ^{\nu _{L}}_{W}}{dTd\varphi}=\mid a_{L}
\mid ^{2}
\frac{G^{2}m_{e}}{\pi ^{2}}\bigg [ g_{L}^{2}+g_{R}^{2}
(1-\frac{T}{E})^{2}-g_{L}g_{R}\frac{m_{e}T}{E_{\nu}^{2}}
\bigg ],
\end{equation}
\begin{equation}
\frac{d\sigma _{em}}{dTd\varphi}=\big ( \frac{\mu}
{\mu _{B}}
\big )^{2}\frac{\alpha ^{2}}{2m_{e}^{2}}\big (
 \frac{1}{T}-\frac{1}{E_{\nu}}\big ),
\end{equation}
\begin{equation}
\frac{d\sigma ^{\nu _{L}}_{int}}{dTd\varphi}=-(\mbox
{\boldmath $\xi$} _{T}\cdot {\bf k})
\big (\frac{\mu}{\mu _{B}}\big )\frac{G\alpha }
{2\sqrt{2}\pi}\frac{1}{m_{e}T}\big [g_{L}+g_{R}(1-
\frac{T}{E_{\nu}})\big ],
\end{equation}
\begin{equation}
\frac{d\sigma ^{\bar \nu _{R}}_{W}}{dTd\varphi}=\mid
a_{R}\mid ^{2}
\frac{G^{2}m_{e}}{\pi ^{2}}\bigg [ g_{R}^{2}+g_{L}'^{2}
(1-\frac{T}{E})^{2}-g'_{L}g_{R}\frac{m_{e}T}{E_{\nu}^{2}}
\bigg ],
\end{equation}
\begin{equation}
\frac{d\sigma ^{\bar \nu _{R}}_{int}}{dTd\varphi}=
(\mbox {\boldmath $\xi$} _{T}\cdot {\bf k})
\big (\frac{\mu}{\mu _{B}}\big )\frac{G\alpha }
{2\sqrt{2}\pi}\frac{1}{m_{e}T}\big [g_{R}+g'_{L}(1-
\frac{T}{E_{\nu}})\big ].
\end{equation}
The  couplings are listed below.
\begin{equation}
g_{L}=\frac{1}{2}+\sin ^{2}\theta _{W},~~~
g_{L}'=-\frac{1}{2}+\sin ^{2}\theta _{W},~~~
g_{R}=\sin ^{2}\theta _{W}.
\end{equation}
The momentum of the recoiling electron is denoted by
${\bf k}$.
Since we have neglected neutrino masses, these cross
sections are common for Majorana  and ZKM neutrinos.
Note that the cross sections (22), (23) and (24)
were considered by Barbieri and Fiorentini [19] and
eqs. (25) and (26) are our new addition.

We have introduced a notation
$\mbox {\boldmath $\xi$} _{T}$,
  the transverse component
of the neutrino spin $<{\bf S}>$  (times 2), namely,
\begin{equation}
\mbox {\boldmath $\xi$} _{T}=2
\left (
\begin{array}{c}
\mid a_{L}a_{R}\mid \cos (\varphi _{L}-\varphi _{R})\\
\mid a_{L}a_{R}\mid \sin (\varphi _{L}-\varphi _{R})\\
0
\end{array}
\right ).
\end{equation}
The existence of
$\mbox {\boldmath $\xi$} _{T}$
in eq. (24) and (26) shows clearly that the azimuthal
distribution of kicked electrons would exhibit asymmetry
with respect to the direction (28).
Barbieri and Fiorentini [19] have calculated
$\varphi _{R}-\varphi _{L}$  as a function of $\mu B$ for
untwisting magnetic field case.   They pointed out that the
asymmetry measurement appears to be within the reach of
observation in future experiments.  The point we would
like to stress here is that the phases
$\varphi _{L}$ and $\varphi _{R}$  are potentially
sensitive to $\phi '$  as we see from the evolution
equation (10) and we anticipate that the axis of the
asymmetry varies from season to season.

Before concluding this section, we present  a formula for
the suppression rate of neutrino flux observed in KII
experiment
\begin{equation}
R({\rm KII}) =
\frac{ \int dE_{\nu}\Phi (E_{\nu}; SSM)\int
_{W}^{T_{\rm max}} dTd\varphi
\frac{d\sigma }{dTd\varphi}}
{\int dE_{\nu}\Phi (E_{\nu}; SSM)\int _{W}^
{T_{\rm max}}dTd\varphi
\frac{d\sigma }{dTd\varphi}(\mid a_{L}\mid ^{2}=1)}.
\end{equation}
Here $\Phi (E_{\nu}; {\rm SSM})$  denotes the neutrino
energy spectrum calculated in the standard solar
model (SSM).  The integration over $T$ is taken from the
threshold energy W of KII detector  up to maximally possible
value $T_{\rm max}=T_{{\rm max}}(E_{\nu})$.

The suppression rate  differs from $\mid a_{L}\mid ^{2}$,
since $(\nu _{\mu L})^{C}$ also contributes to the
detection [20].
If the magnetic moment is on the order of
$(10^{-10}\sim 10^{-11})\mu _{B}$, the electromagnetic
contribution is negligibly small and
this becomes
\begin{equation}
R({\rm KII})=\mid a_{L}\mid ^{2}+0.144 \mid a_{R} \mid ^{2}.
\end{equation}
In passing note that the suppression rate in
 the Homestake
experiment is simply given by
$R({\rm Cl})=\mid a_{L}\mid ^{2}$,
since their detector is blind to $(\nu _{\mu L})^{C}$.

\vskip1.0cm
\begin{center}
{\bf 5. NUMERICAL ANALYSIS
OF THE SOLAR NEUTRINO FLUX
\linebreak
 AND THE AZIMUTHAL ASYMMETRY}
\end{center}

We have undertaken numerical analysis of eq. (10).
The integration is performed from $z=z_{0}$ to
$z=z_{1}$,  where $z_{0}$ and $z_{1}$ are the
entrance and exit coordinates of the magnetic zones
\begin{equation}
z_{0,1}=b\cos (\Delta -\lambda )\pm \sqrt{a^{2}
-b^{2}\sin ^{2} (\Delta -\lambda)}.
\end{equation}
The initial condition is set as $(\tilde{\nu }_{eL},
(\tilde{\nu }_{\mu L})^{C})=(1,0)$  at $z=z_{0}$ and
various parameters are taken as follows:
\begin{equation}
\lambda =\pm 7^{\circ},
{}~~~
\Delta =15 ^{\circ},
{}~~~
a=0.1694R_{\odot},
{}~~~
b=0.8813R_{\odot}.
\end{equation}
The values $\Delta $, $a$, and $b$ were read off
from the  Yoshimura's computer simulation.  The
choice of the latitude of the neutrino's path $\lambda =
7^{\circ}$
 ($-7^{\circ}$) corresponds to the period from August to
October (from February to April).  In these latitude,
the magnetic field strength varies only a little, and
we have assumed in our calculations that $B$ in eq. (11)
is spatially constant.
We take the matter potential [21]
 \begin{equation}
V(z)=0.178R_{\odot}^{-1}\exp [10.84\sqrt{1-
z/R_{\odot}}]{\rm cm} ^{-1}
\end{equation}
in the convective zone.

One might perhaps wonder how it could be that $a+b$ is
greater than the solar radius $R_{\odot}$ as in (32).  The reason
for our choice of the values $a$ and $b$ in (32) is as
follows. Although we said that magnetic field tubes
were  formed as in Fig. 2,   the shape of the tubes is not
  an exact torus according to the Yoshimura's simulation.
Since we are thinking of neutrinos passing through in the
latitude $\lambda=\pm 7^{\circ}$,   we have to take a model
reproducing the magnetic fields at the edge of the  tubes
in lower latitudes.     The choice of the values $a$ and $b$
are taken to realize these realistic magnetic fields
along the neutrino path.

We are basically interested in the $\mu B$-
dependence of the flux of the left-handed electron
 neutrinos $P(\nu _{L})=\mid a_{L}\mid ^{2}=1-
\mid a_{R}\mid ^{2}$, where $a_{L}$ and $a_{R}$ are
the probability amplitudes in (19).
For a better understanding of numerical computation,
we note that at $z=0.85R_{\odot}$, for example,
\begin{equation}
V(z=0.85 R_{\odot})=1.70 \times 10^{-10} {\rm cm}^{-1},
\end{equation}
\begin{equation}
\phi '(z=0.85R_{\odot})=\pm (1.44 \times 10^{-11})
\frac{(2\pi R_{\odot}/X)}{1+5.16\times 10^{-4}
(2\pi R_{\odot}/X)^{2}} {\rm cm}^{-1}.
\end{equation}
This is to be compared with
\begin{equation}
\mu B=2.9 \times 10^{-11} \bigg (
\frac{\mu }{10^{-10}\mu _{B}}\big )
\big (\frac{B}{1 {\rm kG}}\bigg )~{\rm cm}^{-1}.
\end{equation}
In other words, if the twist parameter $X$ is a
 fraction of $R_{\odot}$ and $\mu B$
is of the order of $(1\sim 10)\times 10^{-10}
\mu _{B}{\rm kG}$,
then the seasonal
variation is likely to be seen in $P(\nu _{L})$.

In Fig. 6, $P(\nu _{L})$ is plotted as a function
of $\mu B$, for various twist,
$X=(\pi /3) R_{\odot}$,
$X=\pi R_{\odot}$,
$X=2\pi R_{\odot}$,  and
$X=3\pi R_{\odot}$.
   The solid
(broken) curves correspond to the cases where
neutrinos go through the $7^{\circ}$ latitude in the
northern (southern) hemisphere, that is , around
September (March).
The solid curve denoted by ``toroidal" corresponds to
the $X=\infty$ case.

Characteristic features discussed qualitatively
can be seen in Fig. 6.  When $\phi '(z)$ is negative
as in September, then the off-diagonal element of
the Hamiltonian (3) is more effective and the spin-flip
process occurs more frequently than the opposite case.
In fact the neutrino flux in September is smaller than
that without twist ($X=\infty$).  When $\phi '(z)$ is
positive as in March,  things become opposite, {\it i.e.,}
the suppression of the flux is weakened.  The seasonal
differences are distinct for relatively weak magnetic
field $\mu B=(2\sim 6)\times 10^{-10}\mu _{B}{\rm kG}$.
For stronger magnetic fields, say, $\mu B=(8\sim 10)
\times 10^{-10}\mu _{B}{\rm kG}$,  the seasonal
difference is obscured by the frequent spin-flip
processes and even opposite tendency arises in comparison
with the weak magnetic field.

To confront these numerical computations  with long-term
observational data at Homestake and KII,  we have to
vary the magnetic field strength $B$ in accordance with
the phase of the solar cycle.   In the light of our
poor knowledge on such long-term variation of $B$,
we take here a phenomenological approach.  Namely,
we evaluate $B$  by assuming that it is proportional to
$\sqrt{{\rm sunspots ~number}}$.  In Fig. 7, the
sunspots number in the last two decades are displayed
together with our fit.  At the maximum phase of the
solar cycle,
we input a value $(\mu B)_{{\rm max}}$ , and the 11-year
variation of $B$ is determined by making use of Fig. 7.

In Fig. 8,
we compare our estimation of the
seasonal variation with the $^{37}{\rm Cl}$ data in the
last twenty or so years. Our numerical calculation was
done for two   cases
\begin{description}
\item{(a)}
$X=(\pi /3)R_{\odot}$,~~$(\mu B)_{{\rm max}}=
10\times 10^{-10} \mu _{B}{\rm kG}$,
\item{(b)}
$X=\pi R_{\odot}$,~~$(\mu B)_{{\rm max}}=
9\times 10^{-10} \mu _{B}{\rm kG}$.
\end{description}
Broken lines  connecting predicted values are drawn
only for a guide.  Our calculations show  the seasonal
variations more clearly in the quiet period of the sun
than in the active period. Although the observational
data are afflicted with  large error, it is not unfair
to say that our predicted fluxes    reproduce the data
rather well. One can see an  up-down structure  of the
flux in the Cl data.   Our claim is that
this structure could be interpreted
as a manifestation of the twisting magnetic field.

The KII data are also compared with our numerical
calculations of (29)  in Fig. 9.
Our numerical estimate shows again  an up-down structure
in particular in the quiet period of the sun
({\it i.e.}, 1986-1988).  The available KII data are
those in the maximum phase of the sun and
the expected structure is not so conspicuous.
It is also to be
mentioned that the data are averaged over
7 to 9 months and the up-down   structure, if it were present,
may  have been  washed away by the averaging procedure.
It is more desirable to take data in a larger detector
(like super Kamiokande) without taking average over long
term.

The difference of the azimuthal angle $\varphi _{R}-
\varphi _{L}$ is evaluated in Fig. 10 for the above two
cases, (a) and (b).  Black (white) triangles   are those
around September (March) for the case (a)  and black
(white) circles around September (March) for (b).
  It is obviously seen that the asymmetry
axis swings between two completely different directions.
As far as the asymmetry measurement is possible in a large
volume detector as was remarked by Barbieri and Fiorentini,
the measurement of the seasonal variation in so large
angles should also be feasible.
It is therefore a good proposal in future
 experiments to measure the seasonal variation of
the asymmetry axis.

Finally we add a little comment on the adiabaticity.
In our present calculations we have solved (10) numerically
and there is no reason to worry about the adiabaticity.
It is, however, more instructive to check the validity
of the adiabatic approximation by using our explicit
parametrization  (18) and (33).  The adiabaticity means that
the mixing of states proceeds very slowly during the
passage of neutrinos in a typical precession length.
More explicitly,  the adiabaticity is ensured (near the
resonance point) provided that
\begin{equation}
2(\mu B)^{2}\gg V'(z)+\phi ''(z).
\end{equation}
Numerically the parametrization (33) tells  us
$V'(z)=(2\sim 3)\times 10^{-20} {\rm cm}^{-2}$
for
$0.85<z/R_{\odot}<0.90$.  On the other hand eq. (18)
gives
$\phi ''(z)=(-4\sim 4)\times 10^{-20} {\rm cm}^{-2}$
for
$0.85<z/R_{\odot}<0.90$ and
$X=(\pi/3)R_{\odot}$.
  These values should be compared with
$(\mu B)^{2}=8.41 \times 10^{-22}(\mu /10^{-10}
\mu _{B})^{2}(B/1{\rm kG})^{2}  {\rm cm}^{-2}$.
One can see therefore that for the parameter range of $\mu B$
adopted in our numerical calculations, the adiabaticity
condition  is not well satisfied.  This is of course
consistent with the observation made  previously in
literatures   [10, 12, 17].

\vskip1cm
\begin{center}
{\bf 6. SUMMARY AND DISCUSSION}
\end{center}

In the present paper we have been seeking for a
possibility of making use of the solar neutrino data
to probe the solar magnetic structure.  More specifically
we have examined the effect of twisting magnetic field
to the solar neutrino flux and the azimuthal asymmetry
which could be measured in the super-Kamiokande
experiment.  We have pointed out
that within a reach of observation there should be
semi-annual variation in the solar neutrino flux and
the azimuthal asymmetry axis, provided that the twist
parameter is sizable and $\mu B=(1\sim 10)\times 10^{-10}
\mu _{B}{\rm kG}$.

There have been a lot of attempts of putting upper
bounds on neutrino magnetic moments from both
terrestrial and celestial observations.  The Dirac-type
as well as Majorana-type magnetic moments are constrained
by antineutrino-electron scattering data [22] ($\mid \mu
\mid \leq 4\times 10^{-10}\mu _{B}$),  and stellar energy
loss through neutrino pair emission [23] ($\mid \mu \mid
\leq 1\times 10^{-11}\mu _{B}$).  There is also a
possibility   of deriving a bound from a reactor
experiment [24].  The synthesis of $^{4}{\rm He}$ in
the big   bang gives us a bound for a Dirac- type
magnetic moment [25].   There are several studies of
restricting
the magnetic moment more severely from supernova 1987A [26].
This is, however, still under debate [27]. In any way, the
Majorana-type    moment is not constrained by supernova
1987A.  Considering the ambiguity of the solar magnetic
field strength,  we believe that our choice of the
parameter $\mu B$ is
within a reasonable range.

\vskip1cm
\begin{center}
{\bf ACKNOWLEDGEMENT}
\end{center}
The authors would like to thank M. Ogura for
collaboration in the early stage of the present
work.

\pagebreak
\begin{center}
{\bf REFERENCES}
\end{center}
\begin{description}
\item{[1]} R. Davis, Jr.,   in {\it Neutrino 88},
{\it Proceedings of the XIII International Conference
  on Neutrino and Astrophysics},  Boston, Massachusetts,
1988,  edited by J. Schneps et al. (World Scientific
Pub.,  Singapore, 1989)  p. 518.
\item{[2]} K. Hirata {\it et al.},  Phys. Rev. Lett.
 {\bf 63}, 16 (1989); {\it ibidem} {\bf 65}, 1297;
 1301 (1990).
\item{[3]} P. Anselmann {\it et al.}, Phys. Lett.
{\bf B285}, 390 (1992).
\item{[4]} A.I. Abazov {\it et al.}, Phys. Rev.
Lett. {\bf 67}, 3332 (1991).
\item{[5]} L. Wolfenstein,  Phys. Rev. {\bf D17},
2369 (1978);  {\bf D20},
2634, (1979);
S. P.  Mikheyev and A.Yu.  Smirnov,
Nuovo Cimento {\bf C9},  17 (1986);
Yad. Fiz.  {\bf 42}, (1985) 1441 [Sov. J. Nucl.
Phys. {\bf 42}, 913 (1985)].
\item{[6]}  L.B. Okun, Yad. Fiz. {\bf 44}, 847 (1986)
[Sov. J. Nucl. Phys. {\bf 44}, 546 (1986)] ;
 L.B. Okun, M.B. Voloshin and M.I. Vysotsky, {\it ibidem}
{\bf 91},  754 (1986) [{\bf 44}, 440 (1986)];
Zh. Eksp. Teor. Fiz. {\bf 91}, 754 (1986)
[Sov. Phys. JETP
{\bf 64}, 446 (1986)].
\item{[7]} A. Cisneros, Astrophys. Space Sci. {\bf 10},
87 (1981);  K. Fujikawa and R. Shrock,  Phys. Rev. Lett.
{\bf 45},  963 (1980).
\item{[8]}
C.S. Lim and W. Marciano, Phys. Rev. {\bf D37}, 1368
(1988);
H. Minakata and H. Nunokawa,  Phys. Rev. Lett.
 {\bf 63}, 121 (1989); E.Kh. Akhmedov, Phys. Lett.
{\bf B213}, 64 (1988).
\item{[9]}  S.M. Bilenky and S.T. Petcov, Rev. Mod. Phys.
{\bf 59}, 671 (1987); T.K. Kuo, Rev. Mod. Phys. {\bf 61},
937 (1989);  J. Pluido,  Phys. Reports {\bf 211},
167 (1992).
\item{[10]} J. Vidal and J. Wudka, Phys. Lett. {\bf
B249}, 473 (1990).
\item{[11]} C. Aneziris and J. Schechter, Int. Journ.
Mod. Phys. {\bf A6},  2375 (1991);  Phys. Rev. {\bf D45}
(1992) 1053.
\item{[12]} A. Yu. Smirnov, Phys. Lett. {\bf B260}, 161
(1991); JETP Lett. {\bf 53}, 291 (1991);
 E.Kh. Akhmedov, P.I. Krastev and A.Yu.
Smirnov, Z. Phys. C {\bf 52}, 701 (1991);
A.Yu. Smirnov, in {\it Proceedings  of Joint
International Lepton-Photon Symposium and Europhysics
Conference  on
High Energy Physics } (Geneva, Switzerland, July
1991) p. 648.
\item{[13]} T. Kubota, T. Kurimoto, M. Ogura and
E. Takasugi, Phys. Lett. {\bf B292}, 195 (1992).
\item{[14]} P.D. Mannheim,  Phys. Rev.  {\bf D37},
1935   (1988).
\item{[15]} M.V. Berry, Proc. R.   Soc. London
{\bf A392}, 45 (1984).
\item{[16]} Ya. B. Zel'dovich,  Dok. Akad. Nauk. SSSR
{\bf 86},   505 (1952); E.J. Konopinski and H. Mahmound,
Phy. Rev. {\bf 92}, 1045  (1953).
\item{[17]}
S. Toshev, Phys. Lett.  {\bf B271},  179 (1991).
\item{[18]} H. Yoshimura, Astrophys. Journ.
{\bf 178},  863 (1972); Astrophys. Journ. Suppl.  467,
{\bf  29}: {\it ibidem} {\bf 52}, 363 (1983).
\item{[19]} R. Barbieri and G. Fiorentini,  Nucl. Phys.
{\bf B304}, 909 (1988).
\item{[20]}
C.S. Lim, M. Mori, Y. Oyama and A. Suzuki, Phys. Lett.
{\bf B243}, 389 (1990);
I. Rothstein,  Phys. Rev. {\bf D45}, 2583 (1992);
H.Minakata and H. Nunokawa, Phys. Rev. {\bf D45},
3316 (1992).
\item{[21]}
H. Spruit,  Slolar Phys.  {\bf 34},  277 (1974).
\item{[22]}  W.J. Marciano and Z. Parsa, Annu. Rev.
Nucl. Part. Sci. {\bf 36}, 171 (1986).
\item{[23]}
J. Bernstein, M. Ruderman and G. Feinberg, Phys. Rev.
{\bf 132},  1227 (1963).
\item{[24]}
P. Vogel and J. Engel, Phys. Rev. {\bf D39},
3378 (1989).
\item{[25]}
J.A. Morgan, Phys. Lett.  {\bf 102B}, 247 (1982);
  M. Fukugita and S. Yazaki,  Phys. Rev. {\bf D36},
   3817 (1987).
\item{[26]}
J.M. Lattimer and J. Cooperstein, Phys. Rev. Lett.
{\bf 61}, 23 (1988);  R. Barbieri and R.N. Mohapatra,
Phys. Rev. Lett. {\bf 61}, 27 (1988);  S. Nussinov
and Y.   Rephaeli, Phys. Rev. {\bf D36}, 2278 (1988).
\item{[27]}
M.B. Voloshin,  Phys. Lett. {\bf B209}, 360 (1988).
\end{description}
\pagebreak
\begin{center}
{\bf Figure Captions}
\end{center}
\begin{description}
\item{Fig. 1:}
Schematic pictures of the twist of
toroidal magnetic fields in the convective zone.
The twist is generated by the differential rotations
of the sun and the global convection of the plasma
fluid.  These pictures are given from the work of
Yoshimura [18].
\item{Fig. 2:}
A model of twisting  toroidal magnetic
fields. For simplicity, we assume that magnetic fields
are winding along the tori which are located in the
northern and southern hemispheres.  The parameter
$\lambda $ represents the latitude of the path of
neutrino which is assumed to be created at the center
of the sun.  The latitudes of neutrino paths are restricted
by $\mid \lambda \mid \leq 7.25 ^{\circ}$.
\item{Fig. 3:}
Parametrization of  the twist by $X$
which is the distance along the torus to wind it once.
\item{Fig. 4:}
Twists and the direction of toroidal
magnetic fields.  They change symmetrically, depending on
 the hemisphere and the period.
\item{Fig. 5:}
Description of toroidal magnetic fields.
\item{Fig. 6:}
The $\mu B$ dependence of the left handed neutrino flux
  for various magnitude of the twist parameter
$X=(\pi /3)R_{\odot}$,
$X=\pi R_{\odot}$,
$X=2\pi R_{\odot}$,
$X=3\pi R_{\odot}$.
Solid lines
represent cases where neutrinos pass through the northern
hemisphere,  while broken lines correspond to cases where
neutrinos pass through the southern hemisphere. The curve
denoted by ``toroidal" corresponds to the $X=\infty$ case.
\item{Fig. 7:}
Fit of sunspot numbers which we used to  obtain the time
variation of the magnitude of magnetic fields.
\item{Fig. 8:}
Comparison between our predictions and the Cl-Ar data.
Broken lines are drawn as a  guide. The up-down structure
of our predictions  is the inevitable consequences of the
twist of toroidal magnetic fields.  We have analyzed
two cases:
(a)  $X=(\pi /3)R_{\odot}$,  $(\mu B)_{{\rm max}}=10\times
10^{-10}\mu _{B} {\rm kG}$ and
(b)  $X=\pi R_{\odot}$,  $(\mu B)_{{\rm max}}=9\times
10^{-10}\mu _{B} {\rm kG}$.
\item{Fig.  9:}
Comparison between our predictions and the KII
data.  Our predictions show a structure, but the KII
 data do not.  This is because their data are those
averaged for 7 to    9 months which are too long to see
the structure.  Also their observation are almost in the
maximum phase of the solar magnetic fields where
seasonal difference is minimum.
\item{Fig. 10:}
Theoretical prediction of the difference of the azimuthal
angle $\varphi _{R}-\varphi _{L}$ in the last two decades.
Black (white) triangles  are those around September (March) for
(a) $X=(\pi /3)R_{\odot}, (\mu B)_{{\rm max}}=10\times 10
 ^{-10} \mu _{B}{\rm kG}$.
and black (white) circles those around September (March) for
(b) $X=\pi R_{\odot}, (\mu B)_{{\rm max}}=9\times 10
 ^{-10}\mu _{B}{\rm kG}$.

\end{description}

\end{document}